\documentclass[prb,amsmath,amssymb,superscriptaddress,twocolumn]{revtex4}
\usepackage{graphicx}

\newcommand{\eps}{\epsilon}

\DeclareMathAlphabet{\mathpzc}{OT1}{pzc}{m}{it} 
\begin{document}
\title{Quantum Hall effect in a singly and doubly connected 3D topological insulator}
\author{Oskar Vafek}
\affiliation{National High Magnetic Field Laboratory and Department
of Physics, Florida State University, Tallahassee, Florida 32306,
USA}
\address{}

\date{\today}
\begin{abstract}
The surface states of topological insulators, which behave as
charged massless Dirac fermions, are studied in the presence of a
quantizing uniform magnetic field. Using the method of D.H.
Lee\cite{DHLeePRL2009}, analytical formula satisfied by the energy
spectrum is found for a singly and doubly connected geometry. This
is in turn used to argue that the way to measure the quantized Hall
conductivity is to perform the Laughlin's flux ramping experiment
and measure the charge transferred from the inner to the outer
surface, analogous to the experiment in
Ref.\cite{DolgopolovPhysRevB1992}. Unlike the Hall bar setup used
currently, this has the advantage of being free of the contamination
from the delocalized continuum of the surface edge states. In the
presence of the Zeeman coupling, and/or interaction driven Quantum
Hall ferromagnetism, which translate into the Dirac mass term, the
quantized charge Hall conductivity $\sigma_{xy}=n e^2/h$, with
$n=0,\pm 1,\pm 3,\pm 5\ldots$. Backgating of one of the surfaces
leads to additional Landau level splitting and in this case $n$ can
be any integer.
\end{abstract}
\maketitle

Theoretical prediction of the existence of an odd number of Dirac
cones in the dispersion of the surface states of topological
insulators\cite{FuKaneMelePRL2007,FuKanePRB2007} and subsequent
experimental observation of such unusual surface
states\cite{ZahidHasanNature2008,ZahidHasanScience2009,ZahidHasanNaturePhysics2009,ZXShenScience2009}
has propelled this research field into an active area of condensed
matter physics (for reviews see Refs.
\cite{QiZhangPhysicsToday2010,JoelMoore2010,HasanKaneRevModPhys2010}).
Particularly interesting is the problem of the topological insulator
surface Dirac Fermions in magnetic field. Because the Dirac Fermions
carry definite charge, the magnetic field couples to the orbital
motion. If this motion is constrained to be perpendicular to the
applied field, the Landau level quantization results. However, as
discussed by D.H. Lee\cite{DHLeePRL2009}, since the Dirac Fermions
move on the surface of a 3-dimensional material, in the absence of
magnetic monopoles, i.e for $\nabla\cdot B=0$, it is impossible for
the magnetic field to be {\em everywhere} along the normal of an
oriented surface. Instead, in a typical experimental setting, the
three dimensional material is placed in a uniform external magnetic
field, and only portions of the surface, say the top and the bottom
ones, experience Landau quantization. The Dirac Fermions on the
surfaces tangential to the external magnetic field continue moving
as plane-waves.

In addition, the spin-orbit coupling, which causes the appearance of
the Dirac particles in the first place, makes the Zeeman coupling
different from that in graphene, where Dirac particles also appear
but where the spin-orbit coupling is negligible. Thus, instead of
simply spin splitting the electronic energy levels, the Zeeman term
in topological insulators acts as a Dirac mass. As illustrated in
Fig.\ref{fig:spectrumWithZeeman}, this causes the splitting of the
zeroth Landau level, but the higher Landau levels are not split
unless their guiding center approaches the edge. Rather, at positive
energies they move up and at negative energies they move down. Of
course, because of the Dirac structure, the energy scale associated
with Zeeman splitting $\sim1K\times H[T]$ is much
smaller\cite{DHLeePRL2009} that the spacing between the zeroth and
the first Landau levels, $\sim 200K\times\sqrt{H[T]}$ for realistic
fields. Nevertheless, similar behavior can be expected in the
presence of interaction driven Quantum Hall ferromagnetism where the
"Zeeman" scale can be much larger. While neglecting orbital
coupling, the effects of Zeeman splitting near the boundary between
the portions of the material where the field is perpendicular, and
therefore the Dirac point is gapped, and where it is parallel, and
therefore gapless, was also studied in Ref.\cite{ChuPRB2011}. Some
aspects of Landau quantization in thin films of topological
insulators were also analyzed in Ref.\cite{YangHanPhysRevB2011}. The
analytical results for the energy spectrum obtained in this work are
in agreement with recent numerical lattice band-structure
diagonalization studies in magnetic field for systems with small
withdth\cite{ZhangYY2011arXiv1103.3761Z}.

For a sphere with a finite radius $r$ the Dirac Hamiltonian adopted
to this curved surface in the presence of a uniform applied field
(without Zeeman coupling), was analyzed in an insightful study in
Ref.\cite{DHLeePRL2009}. In the Landau gauge, the eigenenergy is a
function of the azimuthal quantum number $m$ which serves as a
"guiding center" coordinate. For large and positive values of $m$
the spectrum corresponds to nearly doubly degenerate Landau levels,
with the wavefunctions residing near the north and south poles. The
energy splitting is exponentially small in $\sim r/\ell_B$, where
the magnetic length $\ell_B=\sqrt{\hbar c/eB}$. As $m$ approaches
zero, the wavefunctions move towards the equator of the sphere, the
Landau levels are split and merge into the plane-wave-like states
residing near the equator. The spectrum of these states resembles
(finite size) Dirac spectrum. Thus, for the singly connected
topology, the portion of the surface approximately tangential to the
external magnetic field harbors the chiral quantum Hall edge states
{\it together with} the conducting non-chiral surface states, which
disperse as Dirac particles. Since the latter do not
localize\cite{NomuraPRL2008,DHLeePRL2009}, the two terminal
conductance is not quantized even when the Fermi energy lies between
Landau levels. It was proposed\cite{DHLeePRL2009} that the way to
measure the quantized Hall effect is to set up a potential
difference between the electrode placed in the caps of the sphere
and to measure the circulating current on the outer surface.

Experimentally, quantum oscillations originating from the surface
states have been reported by several
groups\cite{YoichiAndoPhysRevB2009,OngPhysRevLett2009,Analytis2010,OngLargeBulkResistivity2011,OngFractions2011}.
In each case (although to varying degree) the signal is contaminated
due to the finite 3D bulk conductivity. In $70nm$ thick strained
films of HgTe, quantum Hall effect has been
reported\cite{BrunePRL2011}, although interestingly the quantized
Hall plateaus appear without the longitudinal resistance $R_{xx}$
reaching zero. So far, no quantization plateaus of the 2D Hall
conductivity has been reported in thick samples. It is important to
ask whether such quantization could be realized in a typical
multi-terminal experimental setup even for a perfectly insulating
bulk. Theoretically, it was argued in
Ref.\cite{ZhangYY2011arXiv1103.3761Z} that for topological
insulators with odd number of Dirac points, the four terminal Hall
conductance should remain quantized even in the presence of scalar
disorder, although the six terminal should not. However, even in the
ballistic limit, such quantization of the four terminal Hall
conductance depends sensitively on the number of non-chiral surface
channels, which may vary between the four electrodes. As such, it is
sensitive to the surface roughness and for typical Fermi
momenta\cite{BrunePRL2011,OngLargeBulkResistivity2011} $k_F\sim
0.2-0.5nm^{-1}$ would require few $nm$ precision in the height of
the sample, as in the molecular beam epitaxy grown films of
Ref.\cite{BrunePRL2011}.

Here we study both the singly and the doubly connected geometry. In
the former case, the standard Hall bar setup
(Fig.\ref{fig:HallaBar}) will not lead to quantization of the Hall
conductivity, unless the sample height is reduced to be much smaller
than the magnetic length. The latter case is illustrated in
Fig.\ref{fig:setup} and argued to be an alternative way to measure
the quantization of $\sigma_{xy}$. The idea is to perform the analog
of the Laughlin thought experiment, experimentally realized in
Ref.\cite{DolgopolovPhysRevB1992}, and to measure the amount of
charge $\Delta Q$ transferred from the inner surface to the outer
surface in response to the change in the magnetic flux $\Delta \Phi$
threading the sample. Then,
\begin{eqnarray}
\sigma_{xy}=-c\frac{\Delta Q}{\Delta \Phi}.
\end{eqnarray}
This setup has the virtue that any interaction driven fractional
quantum Hall effects can also be detected, as shown in the context
of 2DEG heterostructures in Ref.\cite{DolgopolovPhysRevB1992}.

We build on the formalism and findings presented in
Ref.\cite{DHLeePRL2009} and analytically study the energy spectrum
such geometry. We include the Zeeman coupling, as well as a
difference in the gate voltage applied between the top and the
bottom surface of the 3D sample. The former causes the splitting of
the zeroth Landau level, but not the rest of the Landau levels,
while the latter splits all Landau levels. Thus, integer quantum
Hall conductivity, measured in the doubly-connected setup, can take
on any positive or negative integer, including zero.

This paper is organized as follows, in Sec.I the effective
Hamiltonian for the geometry shown in Fig.\ref{fig:setup} is derived
and the boundary conditions at the edges are discussed. In Sec II,
the matching of the wavefunctions at the corners leads to the
equation which determines the condition for the eigenenergies as a
function of the guiding center $k\ell_B$. This equation is shown
analytically to describe a Dirac-like continuum as well as Landau
levels. Its numerical solution determines the behavior of the
discrete levels as the guiding center approaches the continuum. In
Sec III the Hall bar and the Corbino-like geometry are further
compared.

\section{Hamiltonian, eigenstates and the matching conditions}
\begin{figure}[t]
\begin{center}
\includegraphics[width=0.3\textwidth]{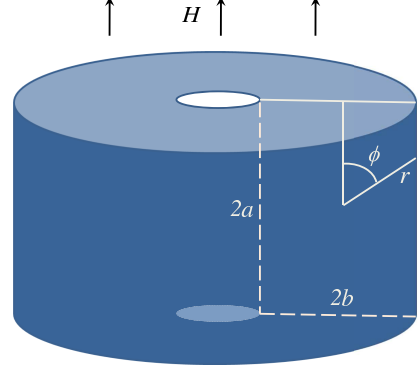}
\end{center}
  \caption{The physical setup proposed for the measurement of $\sigma_{xy}$.
  }\label{fig:setup}
\end{figure}

The specific geometry considered is shown in Fig.\ref{fig:setup}.
The magnetic field is assumed to be perpendicular to the surface of
the hollow cylinder of inner radius $R$, shell thickness $2b$ and
height $2a$. As discussed by D.H. Lee\cite{DHLeePRL2009}, the Dirac
Hamiltonian needs to be written on the surface of this two
dimensional curved space. In the limit of $R\rightarrow \infty$ the
system is equivalent to an infinitely long slab with rectangular
cross-section and periodic boundary conditions along its axis. We
can now use the polar coordinates shown in Fig.\ref{fig:setup} to
describe the surface.

The procedure for determining the Hamiltonian, which follows from
the discussion in Refs.\cite{DHLeePRL2009,Pnueli1994}, is shown the
Appendix. The result is
\begin{widetext}
\begin{eqnarray}\label{eq:startingHamiltonian}
H&=&\hbar v_F
\left[\sigma_1\left(\frac{1}{i}\frac{\partial}{\partial
x}-\frac{a}{\ell^2_B}\tan\phi\right)+\sigma_2\frac{1}{i}\frac{\partial}{\partial
a\tan\phi}\right]+h_z\sigma_3,\;\;\;\;\; -\phi_0<\phi<\phi_0\nonumber\\
&=&\hbar v_F \left[\sigma_1\left(\frac{1}{i}\frac{\partial}{\partial
x}-\frac{b}{\ell^2_B}\right)-\sigma_2\frac{1}{i}\frac{\partial}{\partial
b\cot\phi}\right]-h_z\sigma_1,\;\;\;\;\; \phi_0<\phi<\pi-\phi_0\nonumber\\
&=&\hbar v_F \left[\sigma_1\left(\frac{1}{i}\frac{\partial}{\partial
x}+\frac{a}{\ell^2_B}\tan\phi\right)+\sigma_2\frac{1}{i}\frac{\partial}{\partial
a\tan\phi}\right]+h_z\sigma_3+V_g,\;\;\;\;\; \pi-\phi_0<\phi<\pi+\phi_0\nonumber\\
&=&\hbar v_F \left[\sigma_1\left(\frac{1}{i}\frac{\partial}{\partial
x}+\frac{b}{\ell^2_B}\right)-\sigma_2\frac{1}{i}\frac{\partial}{\partial
b\cot\phi}\right]+h_z\sigma_1,\;\;\;\;\;
\pi+\phi_0<\phi<2\pi-\phi_0.
\end{eqnarray}
\end{widetext}
where periodic boundary conditions for $x$, which is allowed to get
arbitrarily large, are assumed. Obviously, the wavefunctions
separate and correspond to planewaves with wavector $k$ in the
$x$-coordinate. This will serve as the analog of the guiding center
coordinate. The term proportional to $h_z$ represents the Zeeman
coupling. A mean-field description of any possible quantum Hall
ferromagnetism would have a different value of $h_z$ in the first
and the third region than in the second and fourth. As usual, the
magnetic length is $\ell_B=\sqrt{\hbar c/eB}$.

We can now find the eigenfunctions at a fixed energy $E$ and a fixed
$k$ for each segment. This is done in detail in the Appendix. For
the 1$^{st}$ and 3$^{rd}$ segment the Dirac eigenspinors can be
written in a closed form in terms of the parabolic cylinder
functions, $D_{\nu}(z)$, with indices $\nu$ of the upper and the
lower components differing by $1$. For the 2$^{nd}$ and 4$^{th}$
segment, the eigenspinors are plane-waves. In what follows, we use
dimensionless lengths and energy scales defined as
\begin{eqnarray} \alpha
&=&\frac{a}{\ell_B},\;\beta=\frac{b}{\ell_B},\;\kappa=k\ell_B\\
\eps &=&\frac{E}{\hbar v_F/\ell_B},\;\eta_z=\frac{h_z}{\hbar
v_F/\ell_B},\;\nu_g=\frac{V_g}{\hbar v_F/\ell_B}.
\end{eqnarray}

The matching of the wavefunction is discussed in detail below. While
in general 4-parameter family of self-adjoined extensions is
allowed, it is argued that for sharp edges, continuity of the
wavefunctions should be imposed.

\subsection{Matching conditions}

In order to completely specify the behavior of the wavefunctions,
the Hamiltonian (\ref{eq:startingHamiltonian}) must be supplemented
by boundary conditions at the points where the horizontal and the
vertical surfaces meet. We will assume that the corners are sharp,
meaning that the lengthscale associated with the corner curvature is
much smaller than the magnetic length $\ell_B$.

If we require that the Hamiltonian is a self-adjoint operator we
require that any two spinor wavefunctions $\psi_{1}$ and $\psi_{2}$
satisfy
\begin{eqnarray}\label{eq:hermiticity}
\psi^{\dagger}_{1L}(\phi_0)\sigma_2\psi_{2L}(\phi_0)&=&
\psi^{\dagger}_{1R}(\phi_0)\sigma_2\psi_{2R}(\phi_0),
\end{eqnarray}
where the subscripts $L$ and $R$ refer to the direction of approach
of the boundary, i.e. left or right\cite{StoneGoldbartBook}. The
most general linear homogeneous boundary condition imposed on
$\psi_2$ is
\begin{eqnarray}
\psi_{2L}&=&M\psi_{2R},
\end{eqnarray}
where $M$ is a $2$ by $2$ matrix. In order to satisfy
(\ref{eq:hermiticity}) we then must have
\begin{eqnarray}
\psi^{\dagger}_{1L}(\phi_0)\sigma_2M\psi_{2R}(\phi_0)&=&
\psi^{\dagger}_{1R}(\phi_0)\sigma_2\psi_{2R}(\phi_0).
\end{eqnarray}
This must hold for arbitrary $\psi_{2R}$ and therefore the boundary
condition on $\psi_1$ is
\begin{eqnarray}
\psi_{1R}&=&\sigma_2 M^{\dagger}\sigma_2\psi_{1L}.
\end{eqnarray}
Since we require that the domains of $H$ and $H^{\dagger}$ coincide,
the above must also be the boundary conditions on $\psi_2$.
Therefore, $M$ must satisfy
\begin{eqnarray}
M^{-1}&=&\sigma_{2}M^{\dagger}\sigma_2.
\end{eqnarray}
Taking the determinant of both sides gives
\begin{eqnarray}
\frac{1}{\det M}&=&\det M^*\Rightarrow \det M=e^{i\chi}.
\end{eqnarray}
Using the above relations we finally find that
\begin{eqnarray}
M&=&e^{\frac{i}{2}\chi}\left(
\begin{array}{cc}
W+Z & X-Y\\
X+Y & W-Z
\end{array}
\right),\\
1&=&W^2-Z^2-X^2+Y^2.
\end{eqnarray}
That means that the self-adjoint constraint leaves us with $4$
independent real paraments which determine the boundary conditions
at the corners. The von Neumann-Weyl deficiency
indices\cite{StoneGoldbartBook} are therefore $(2,2)$

To determine these conditions we regularize the corners as small
circles with the vertical and horizontal surface lines being
tangential to the circle. Eventually, we are interested in taking
the limit of the radius of the circle, $r_0$, to zero. Note that in
the limit $r_0\ll \ell_B$ the vector potential does not depend on
the position along the circle. As a result, the solution to the
Dirac equation along the circle is a plane wave. The ratio of the
two components of the Dirac spinor of the planewave solution is
independent of the position, meaning that $X=Y=Z=0$. At finite
energies the wave vector must be finite and therefore in the limit
of $r_0\rightarrow 0$ the phase does not advance, i.e. $\chi=0$.
This argument leads us to the requirement that for sharp corners,
the Dirac spinors are continuous, or
\begin{eqnarray}
M=1_2.
\end{eqnarray}

In addition, we are going to take the limit $b\gg \ell_B$ and focus
on the solutions with the "guiding center" $\kappa=k\ell_B$ either
$\approx b/\ell=\beta$ or $\approx -\beta$. In the former case,
$\kappa\approx \beta$, we can require that the wavefunction vanishes
for $\tan\phi\rightarrow -\infty$ on the top surface and for
$\tan\phi\rightarrow \infty$ on the bottom surface. Similarly, for
$\kappa\approx -\beta$, we can require that the wavefunction
vanishes for $\tan\phi\rightarrow \infty$ on the top surface and for
$\tan\phi\rightarrow -\infty$ on the bottom surface.

\section{The energy spectrum}
\begin{figure}[t]
\begin{center}
\includegraphics[width=0.5\textwidth]{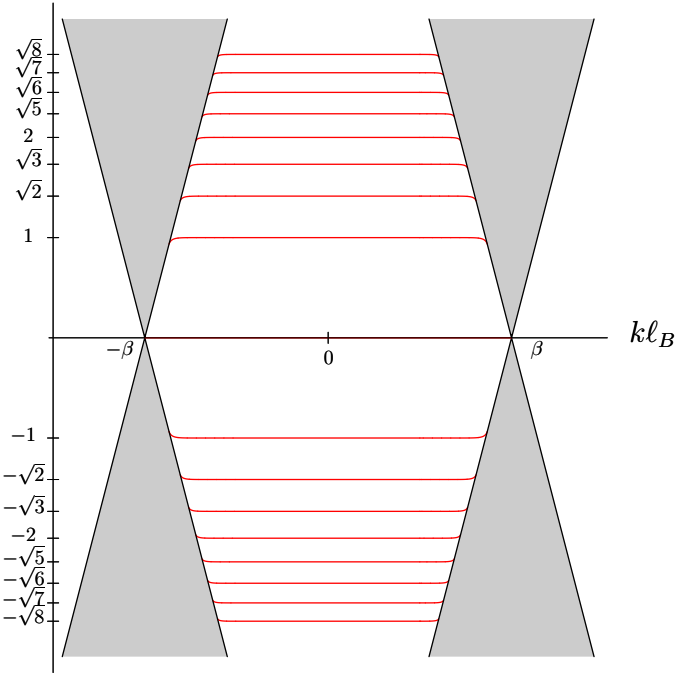}
\end{center}\label{fig:spectrumNoZeeman}
  \caption{Energy spectrum for $h_Z=0$ and $a\rightarrow\infty$. In addition $\beta=b/\ell_B\gg 1$.
  The shaded cones represent the Dirac continuum physically located at the inner (left) and the outer (right)
  surfaces of the doubly connected sample. Each Landau level is doubly degenerate and at energy $\sqrt{2n}\hbar v_F/\ell_B$.
  To understand how the discrete Landau levels merge into the continuum, see Fig.\ref{fig:spectrumFS}}
\end{figure}
The energy spectrum at each $k$ is determined from the matching
conditions on the wavefunctions discussed above. Near each edge,
these take the form of four linear equations in four unknowns
(\ref{eq:matchingOuterSurface}) and (\ref{eq:matchingInnerSurface}),
with energy and $k$ dependent coefficients. The technical details
are presented in the Appendix. Here we present the analysis of the
energy spectrum which results from these equations.

 The non-trivial solution to the
Eq.(\ref{eq:matchingOuterSurface}) exists provided that
\begin{eqnarray}\label{eq:spectrumOuterSurface}
&&\left[(\omega_-+i\theta_-)\mathcal{B}_{+}-\eps
\mathcal{A}_{+}\right]\left[(\omega_--i\theta_-)\mathcal{A}_{g+}-\eps
\mathcal{B}_{g+}\right]e^{2i\theta_{-}\alpha}
=\nonumber\\
&&\left[(\omega_--i\theta_-)\mathcal{B}_{+}-\eps
\mathcal{A}_{+}\right]\left[(\omega_-+i\theta_-)\mathcal{A}_{g+}-\eps
\mathcal{B}_{g+}\right]e^{-2i\theta_{-}\alpha}.\nonumber\\
\end{eqnarray}
The functions used in the above equation are derived in the
Appendix. The above equation determines the energy spectrum for
$\kappa\approx \beta$, i.e. near the outer surface. If we are
interested in the finite size effects, we have to keep
$\alpha=a/\ell_B$ finite. This is useful to understand the
degeneracy of each energy level caused by the tunneling across the
vertical edge as well as mesoscopic transport effects discussed in
the next section. On the other hand, in the thermodynamic limit
$\alpha\rightarrow \infty$.

In order to solve Eq.(\ref{eq:spectrumOuterSurface}) for
$\eps=E/(v_F/\ell_B)$ we need to consider two cases.

For $\theta^2_{-}>0$ the left hand side of
Eq.(\ref{eq:spectrumOuterSurface}) is the complex conjugate of the
right hand side. Therefore, we are looking for roots of a purely
imaginary function. Letting
\begin{eqnarray}
\mathcal{S}=\left[(\omega_-+i\theta_-)\mathcal{B}_{+}-\eps
\mathcal{A}_{+}\right]\left[(\omega_--i\theta_-)\mathcal{A}_{g+}-\eps
\mathcal{B}_{g+}\right],
\end{eqnarray}
we find that (\ref{eq:spectrumOuterSurface}) requires
\begin{eqnarray}
\tan\left(2\theta_-\alpha\right)=-\frac{\Im m\mathcal{S}}{\Re e
\mathcal{S}}.
\end{eqnarray}
Note that the right hand side of the above equation does not depend
on $\alpha$ while the left hand side has a single pole every time
$2\theta_-\alpha=m\pi$ where $m=0,\pm1,\pm2\ldots$. For
$\alpha\rightarrow \infty$ the spacing between the poles approaches
zero and the left hand side changes from $-\infty$ to $\infty$ for
$m\pi/(2\alpha)<\theta_-<(m+1)\pi/(2\alpha)$. Since the right hand
side is a function of $\eps$ which changes on scales much larger
than $1/\alpha$ as $\alpha\rightarrow\infty$, we have at least one
positive and one negative eigenenergy solution for each interval
$m\pi/(2\alpha)<\theta_-<(m+1)\pi/(2\alpha)$. This proves that for
$$\eps^2>(\kappa-\beta-\eta_z)^2$$
the energy spectrum forms a continuum as $\alpha\rightarrow\infty$.

\begin{figure}[t]
\begin{center}
\includegraphics[width=0.5\textwidth]{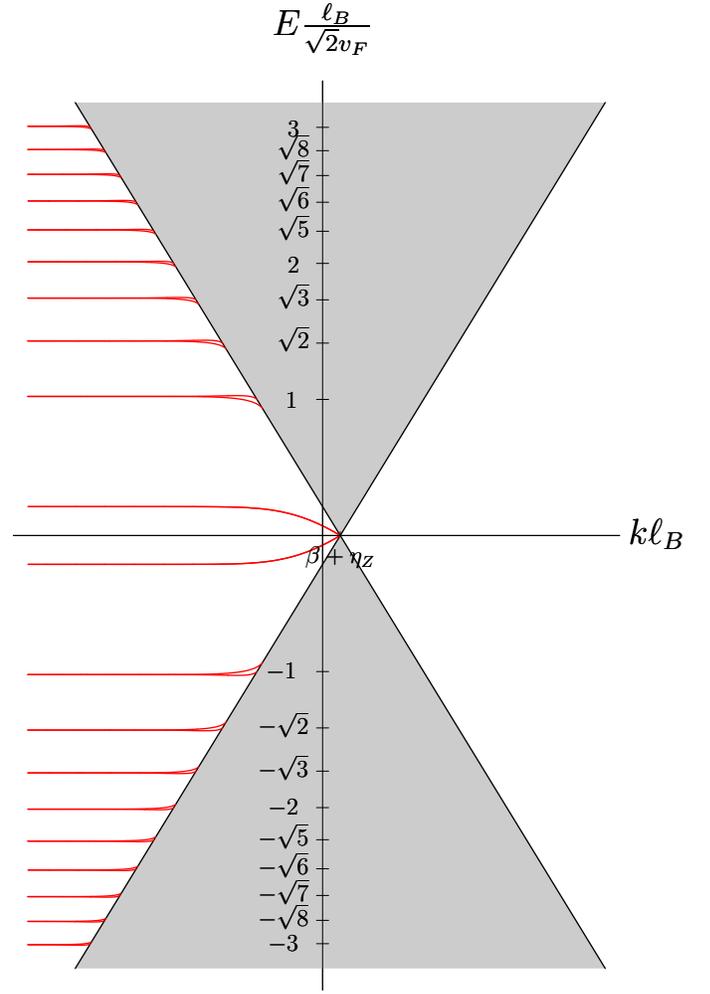}
\end{center}\label{fig:spectrumWithZeeman}
  \caption{Energy spectrum for $h_Z\neq0$ ($\eta_z=0.3$) and $a\rightarrow\infty$ near the outer surface. In addition $\beta=b/\ell_B\gg 1$.
  The shaded cones represent the Dirac continuum physically located at the outer surface. The degeneracy of the
  Landau levels is split as $\kappa=k\ell_B$ approaches the continuum. The double degeneracy of the zeroth Landau level is lifted,
  with a gap that varies linearly in $h_Z$. The double degeneracy of the higher Landau levels is split near the continuum, but in the thermodynamic limit
  remains intact for $\kappa\ll k\ell_B$ with the energies $\pm (\hbar v_F/\ell)\sqrt{2n+\eta^2_z}$, $n=0,1,2,\ldots$.
  The spectrum near the inner surface can be obtained
  simply by taking $\kappa\rightarrow -\kappa$, i.e. by mirror reflecting the above picture around $k\ell_B=0$.}
\end{figure}

For $\theta^2_{-}<0$, the left hand side vanishes in the limit
$\alpha\rightarrow \infty$ and the spectrum is determined by
\begin{eqnarray}
\left[(\omega_-+|\theta_-|)\mathcal{B}_{+}-\eps
\mathcal{A}_{+}\right]\left[(\omega_--|\theta_-|)\mathcal{A}_{g+}-\eps
\mathcal{B}_{g+}\right]=0.\nonumber\\
\end{eqnarray}
For $\kappa-\beta\ll-1$, we can solve these equations within
exponential accuracy by noting that for non-negative integer, the
parabolic cylinder functions entering $\mathcal{A}_{+}$ and
$\mathcal{B}_{+}$ satisfy
\begin{eqnarray}
D_{\nu}(z)&=&2^{-\nu/2}e^{-\frac{z^2}{4}}H_{\nu}\left(\frac{z}{\sqrt{2}}\right),
\;\mbox{for}\; \nu=0,1,2,\ldots,\nonumber\\
\end{eqnarray}
where $H_n(z)$ is Hermite polynomial. If the $\nu$ in $D_{\nu}(z)$
deviates from non-negative integer, the function diverges
exponentially at large negative $z$. Therefore, as long as
$\kappa-\beta\ll-1$, the equation
$$
\left[(\omega_-+|\theta_-|)\mathcal{B}_{+}-\eps
\mathcal{A}_{+}\right]=0
$$
is solved for
\begin{eqnarray}
\eps &=&\pm\sqrt{2n+\eta^2_z},\;\;\; n=1,2,3,\ldots,\nonumber\\
\eps &=&-\eta_z.
\end{eqnarray}
Similarly, as long as $\kappa-\beta\ll-1$ the solutions to
$$\left[(\omega_--|\theta_-|)\mathcal{A}_{g+}-\eps
\mathcal{B}_{g+}\right]=0$$ are
\begin{eqnarray}
\eps &=&\nu_g\pm\sqrt{2n+\eta^2_z},\;\;\; n=1,2,3,\ldots,\nonumber\\
\eps &=&\nu_g+\eta_z.
\end{eqnarray}
For $\kappa\approx\beta$ the equations can be easily solved
numerically and the dispersion of the Landau levels in the vicinity
of the Dirac continuum can be determined. The solution for
$\eta_z=\nu_g=0$ is shown in Fig.\ref{fig:spectrumNoZeeman} for both
edges. In this case all Landau levels are doubly degenerate and the
$\sigma_{xy}$ sequence is $\frac{e^2}{h}\times (\mbox{odd
integer})$. Also, note the downward dispersion of the Landau levels
as they approach the continuum. For $\eta_z=0.3$, $\nu_g=0$ the
energy spectrum is shown in Fig. \ref{fig:spectrumWithZeeman} for
outer the edge. The lowest Landau level is now split linearly in
$h_z$ giving rise to the possibility that if the Fermi energy lies
in this gap $\sigma_{xy}=0$. The higher Landau levels are doubly
degenerate, but split as they approach the edge continuum. Therefore
for $\nu_g=0$, the allowed values for $\sigma_{xy}=n e^2/h$, are
$n=0,\pm 1,\pm 3,\pm 5\ldots$. Finally, for $\nu_g\neq0$ all
possible Landau level (double) degeneracies are lifted, and as a
result $n=0,\pm 1,\pm 2,\pm 3\ldots$.

Similarly, the non-trivial solution to the
Eq.(\ref{eq:matchingInnerSurface}) exists provided that
\begin{eqnarray}
&&\left[(\omega_++i\theta_+)\mathcal{B}_{-}-\eps
\mathcal{A}_{-}\right]\left[(\omega_+-i\theta_+)\mathcal{A}_{g-}-\eps
\mathcal{B}_{g-}\right]e^{-2i\theta_{+}\alpha}
=\nonumber\\
&&\left[(\omega_+-i\theta_+)\mathcal{B}_{-}-\eps
\mathcal{A}_{-}\right]\left[(\omega_++i\theta_+)\mathcal{A}_{g-}-\eps
\mathcal{B}_{g-}\right]e^{2i\theta_{+}\alpha}\nonumber\\
\end{eqnarray}
The above equation determines the energy spectrum for $\kappa\approx
-\beta$, i.e. near the inner surface. The analysis of this equation
proceeds in the same way as for Eq.(\ref{eq:spectrumOuterSurface}).
The continuum appears for
$$
\eps^2>(\kappa+\beta+\eta_z)^2.
$$
Furthermore, the equations for $\eps^2<(\kappa+\beta+\eta_z)^2$ can
be shown to be the same as (\ref{eq:spectrumOuterSurface}) under the
transformation $\kappa\rightarrow -\kappa$. The spectrum near the
inner surface can therefore be determined simply by mirror
reflecting the spectrum near outer surface about $\kappa=0$.

\section{Discussion of Hall bar vs. "Corbino" geometry}
\begin{figure}[t]
\begin{center}
\includegraphics[width=0.5\textwidth]{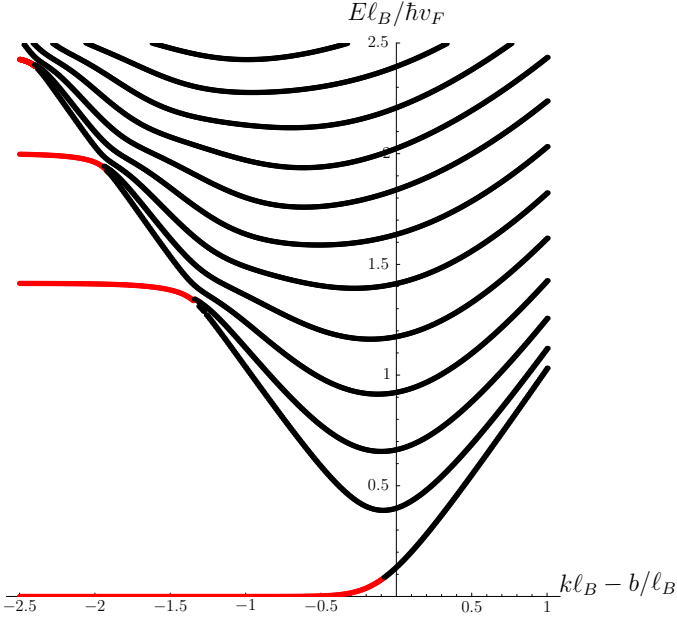}
\end{center}\label{fig:spectrumFS}
  \caption{The positive energy spectrum for $h_Z=0$ and $\alpha=a/\ell_B=5$ near the outer surface. In addition $\beta=b/\ell_B\gg
  1$.  Note the presence of the discrete non-chiral modes, which merge into continuum for $\alpha\rightarrow \infty$.
  The spacing between the levels at $k\ell_B=b/\ell_B$ scales as $\sim 1/a$
  while the spacing between the (Dirac) Landau levels for $k\ell_B\ll b/\ell_B$ scales as $\sim 1/\ell_B$. Therefore,
  in order to observe quantized Hall conductivity in the Hall bar geometry the condition $a\ll \ell_B$ should be satisfied.}
\end{figure}

\begin{figure}[t]
\begin{center}$
\begin{array}{cc}
\includegraphics[width=0.25\textwidth]{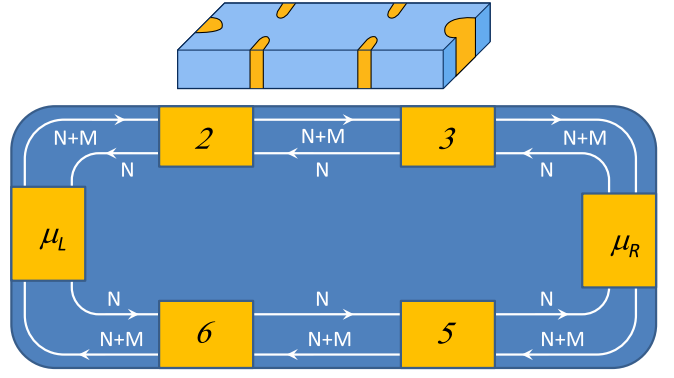} \\
\includegraphics[width=0.48\textwidth]{fig5b.epsi}
\end{array}$
\end{center}
  \caption{(Top panel) A schematic for a Hall bar setup for the 3D topological insulator. The applied magnetic field is perpendicular to the top
  surface. The applied magnetic field is perpendicular to the top surface; six contacts are also marked. (Bottom panel) The
  corresponding edge/surface state structure. There are $M$ chiral
  modes coming from the top and the bottom surface Hall droplet. In
  the absence of the bottom gate $M$ can be any odd integer (including zero if Zeeman field is included).
  In addition, there are $N$ non-chiral modes coming from the
  surfaces parallel to the external field.
  }\label{fig:HallaBar}
\end{figure}

Currently, the experimental geometry used to measure the Hall
conductivity in the quantum Hall regime of 3D topological insulators
is the Hall bar geometry sketched in Fig.\ref{fig:HallaBar}. No
plateau quantization of $\sigma_{xy}$ has been observed. While part
of the reason for this is finite 3D bulk conductivity, we wish to
argue here that even if the system was insulating in the bulk the
presence of the non-chiral surface modes will spoil the quantization
of $\sigma_{xy}$. One way to avoid such contamination would be to
reduce the sample height $\ll \ell_B$. The second way, would be to
use "Corbino" geometry shown in Fig.\ref{fig:setup}, to ramp up the
flux through the hollow region and to measure the charge transferred
between the inner and the out surfaces.

For the Hall bar geometry in the ballistic limit, the Hall
conductance, as well as the longitudinal conductance, are easily
obtained within the Landauer-Buttiker formalism. Assuming that the
contacts $2$, $3$, $5$ and $6$ float to the average chemical
potential of the modes which enter them, for $M$-chiral modes and
$N$-non-chiral modes with perfect transmission the Hall conductance,
measured between contacts $2$ and $6$, is
\begin{eqnarray}
G_{Hall}&=&\frac{e^2}{h}\left(M+N+\frac{N^2}{M}\right).
\end{eqnarray}
The longitudinal conductance measured between $2$ and $3$ is
\begin{eqnarray}
G&=&\frac{e^2}{h}\left(2N+M+\frac{M^2}{N}+\frac{M^2}{N+M}\right).
\end{eqnarray}
For finite $N$, the Hall conductance here is generally not
quantized. In the thermodynamic limit, $N\gg M$ and the Hall
resistance is small ($\sim \frac{h}{e^2}M/N^2$). These results were
also obtained in Ref.\cite{ZhangYY2011arXiv1103.3761Z}.

The physical reason for the non-quantization for large sample height
is quite clear. The presence of the large number of non-chiral
surface states tends to make the local potential on {\em each} of
the surfaces close to $(\mu_L+\mu_R)/2$, as opposed to $\mu_L$ on
one, and $\mu_R$ on the other, as is the case in the absence of
non-chiral modes. In the presence of disorder, in the form of scalar
potential in the Dirac equation, but in the absence of
electron-electron interactions, the states have been argued to
remain delocalized \cite{BardarsonPRL2007,NomuraPRL2008} and the
conductivity diverges as temperature $T\rightarrow 0$. Therefore in
this limit there will be no voltage drop along the surfaces, all of
which will appear near the contacts. The local potential near the
two surfaces will still tend to $(\mu_L+\mu_R)/2$ and no
quantization of $G_{Hall}$ will occur. Including electron-electron
interactions and scalar disorder, it has been argued in
Ref.\cite{OstrovskyPRL2010} that the states remain delocalized, but
that at $T=0$ the conductivity flows to a finite value. Therefore,
the potential will drop along the surfaces, the modes will
equilibrate but no quantization will occur. The two terminal
conductance will also not be quantized\cite{DHLeePRL2009}. The way
to achieve quantization in the Hall bar geometry is therefore to
eliminate the side surface modes altogether, which can be achieved
through finite size effects by reducing the height of the sample
below $\ell_B$.

A four terminal setup shown in Fig.6a of
Ref.\cite{ZhangYY2011arXiv1103.3761Z} has been argued to lead to
quantized Hall conductance $Me^2/h$ even in the presence of
disorder. However, this result was obtained assuming that the number
of non-chiral channels is exactly $N$ for each of the four segments
of the four terminal setup. Unlike the number of chiral channels
$M$, for a thick sample the number of non-chiral channels can vary
from segment to segment. In the four terminal setup, we should
therefore consider $N_1$ channels between $\mu_L$ and $\mu_2$, $N_2$
between $\mu_2$ and $\mu_R$, $N_3$ between $\mu_R$ and $\mu_4$ and
$N_4$ between $\mu_4$ and $\mu_L$. In the ideal case of perfect
transmission, we find
$I_4-I_2=\frac{e^2}{h}\left(2M+N_1-N_2+N_3-N_4\right)(\mu_L-\mu_2)$.
Therefore, unless $N_1-N_2=N_4-N_3$, the Hall conductance defined
this way is not properly given by the number of chiral channels. As
mentioned in the introduction, for typical Fermi
momentum\cite{BrunePRL2011,OngLargeBulkResistivity2011} $k_F\sim
0.2-0.5nm^{-1}$, this would require a few $nm$ precision in the
height of the sample. It seems that, in the least, such sensitivity
to surface roughness would have to be eliminated in practice.

In the Corbino geometry, the increase of the flux will transfer the
charge from the inner to the outer surface, which can then measured.
In the thermodynamic limit, this quantization is robust to the
presence non-chiral states. Such measurement of $\sigma_{xy}$ has
been performed in 2DEGs (Ref.\cite{DolgopolovPhysRevB1992}) where
both integer and fractional quantizations have been detected, and
should be feasible in the 3D topological insulators in the quantum
Hall regime.

\begin{acknowledgments}
I wish to thank Profs. Kun Yang, Nick Bonesteel and Dr. Bitan Roy
for discussions and Profs. Boebinger and N.P. Ong for encouragement
to write up this work. The work was supported by NSF CAREER award
under Grant No. DMR-0955561. After this work was completed a
preprint studying related problem appeared on the
arXiv\cite{Sitte2011}. Where the two overlap, the results are
compatible.
\end{acknowledgments}
\appendix
\section{Derivation of the Hamiltonian, eigenstates and the matching conditions}
In this appendix we present the detailed steps which lead to the
Eq.(\ref{eq:startingHamiltonian}) as well as to the matching
conditions which lead to the equation for the energy spectrum. We
use the parametrization shown in Fig.\ref{fig:setup}.

\subsection{Top horizontal surface ($-\phi_0<\phi<\phi_0$)}
For the top horizontal surface, $r(\phi)=a/\cos\phi$ and
$-\phi_0<\phi<\phi_0$ where $\tan\phi_0=b/a$. For $\phi$ in this
range the metric for the surface is
\begin{eqnarray}
ds^2&=&dx_1^2+dx_2^2=dx^2+\left[r^2(\phi)+\left(\frac{dr(\phi)}{d\phi}\right)^2\right]d\phi^2\nonumber\\
&=&dx^2+\left(d a\tan\phi\right)^2.
\end{eqnarray}
In the Landau gauge of choice here, the two components of the vector
potential are
\begin{eqnarray}
A_1&=&Ba\tan\phi,\;\; A_2=0.
\end{eqnarray}

The physical spin is related to the Dirac Pauli matrices
by\cite{DHLeePRL2009}
\begin{eqnarray}
\vec{s}=\hat{n}\times \vec{\sigma}
\end{eqnarray}
where $\hat{n}$ is the normal to the surface. Using
$\hat{n}=\hat{z}$ and $s_1s_2=is_3$, we obtain
\begin{eqnarray}
s_1&=&-\sigma_2,\;\; s_2=\sigma_1,\;\; s_3=\sigma_3.
\end{eqnarray}
Thus, for $-\phi_0<\phi<\phi_0$ the Hamiltonian is
\begin{eqnarray}
H&=&\hbar v_F
\left[\sigma_1\left(\frac{1}{i}\frac{\partial}{\partial
x}-\frac{a}{\ell^2_B}\tan\phi\right)+\sigma_2\frac{1}{i}\frac{\partial}{\partial
a\tan\phi}\right]+h_z\sigma_3.\nonumber\\
\end{eqnarray}
$\ell_B=\sqrt{\frac{\hbar c}{eB}}$. This is clearly separable, and
the eigenfunctions are planewaves in the $x-$ direction. Its
wavevector is set to be $k$. Letting
$\rho=\sqrt{2}\left(\frac{a}{\ell_B}\tan\phi-k\ell_B\right)$ leads
to
\begin{eqnarray}
\left(\begin{array}{cc} h_z-E & -\sqrt{2}\frac{\hbar v_F}{\ell_B}\left(\frac{\partial}{\partial \rho}+\frac{1}{2}\rho\right)\\
\sqrt{2}\frac{\hbar v_F}{\ell_B}\left(\frac{\partial}{\partial
\rho}-\frac{1}{2}\rho\right) & -h_z-E\end{array}\right)
\left(\begin{array}{c}u_{k}(\phi)\\v_{k}(\phi) \end{array}\right)=0.
\end{eqnarray}
The generic solution to these two coupled first order differential
equations is
\begin{eqnarray}\label{eq:solutionTopSurface}
\left(\begin{array}{c}u_{k}(\phi)\\v_{k}(\phi)
\end{array}\right)=c_{1,L}\left(\begin{array}{c}\alpha_1D_{\nu}(\rho)\\ \beta_1D_{\nu+1}(\rho)
\end{array}\right)+c_{1,R}\left(\begin{array}{c}\alpha_2D_{\nu}(-\rho)\\ \beta_2D_{\nu+1}(-\rho)
\end{array}\right),
\end{eqnarray}
where $D_{\nu}(\rho)$ is the parabolic cylinder function
\cite{MerzbacherQM1998}
\begin{eqnarray}
D_{\nu}(\rho)&=&2^{\nu/2}e^{-\rho^2/4}\left[
\frac{\Gamma[\frac{1}{2}]}{\Gamma[(1-\nu)/2]}
{_1}F_1\left(-\frac{\nu}{2};\frac{1}{2};\frac{\rho^2}{2}\right)\right.\nonumber\\
&&\left.\frac{\rho}{\sqrt{2}}\frac{\Gamma[-\frac{1}{2}]}{\Gamma[-\nu/2]}
{_1}F_1\left(\frac{1-\nu}{2};\frac{3}{2};\frac{\rho^2}{2}\right)
\right]
\end{eqnarray}
and ${_1}F_1$ is the confluent hypergeometric function
\begin{eqnarray}
{_1}F_1\left(a;b;\rho\right)=1+\frac{a}{c}\frac{\rho}{1!}+\frac{a(a+1)}{c(c+1)}\frac{\rho^2}{2!}+\ldots.
\end{eqnarray}
Unless $\nu$ is a non-negative integer, the two solutions in the
Eq.(\ref{eq:solutionTopSurface}) are linearly independent and
$D_{\nu}(\rho)$ diverges as $\rho\rightarrow -\infty$. These
functions satisfy the relations
\begin{eqnarray}
\left(\frac{\partial}{\partial
\rho}+\frac{1}{2}\rho\right)D_{\nu}(\rho)&=&\nu D_{\nu-1}(\rho),\\
\left(\frac{\partial}{\partial
\rho}-\frac{1}{2}\rho\right)D_{\nu}(\rho)&=&-D_{\nu+1}(\rho).
\end{eqnarray}

Since we are interested in taking the limit $b\gg \ell_B$, the
solution near the outer surface must satisfy vanishing boundary
condition as it moves closer to the 2D "bulk". This means that near
the outer surface $c_{1,L}=0$ and
\begin{eqnarray}\label{eq:solutionTopOuterSurface}
\left(\begin{array}{c}u_{k}(\phi)\\v_{k}(\phi)
\end{array}\right)=c_{1,R}\left(\begin{array}{c}(\eps+\eta_z)D_{\frac{1}{2}(\eps^2-\eta_z^2)-1}\left(\sqrt{2}\left(\kappa-\alpha\tan\phi\right)\right)\\
\sqrt{2}
D_{\frac{1}{2}(\eps^2-\eta_z^2)}\left(\sqrt{2}\left(\kappa-\alpha\tan\phi\right)\right)
\end{array}\right),
\end{eqnarray}
where the dimensionless lengths and energy scales are
$\alpha=\frac{a}{\ell_B}$, $\beta=\frac{b}{\ell_B}$,
$\kappa=k\ell_B$, $\eps =\frac{E}{\hbar v_F/\ell_B}$,
$\eta_z=\frac{h_z}{\hbar v_F/\ell_B}$ and $\nu_g=\frac{V_g}{\hbar
v_F/\ell_B}$.

Similarly, near the inner surface $c_{1,R}=0$ and the solution has
the form
\begin{eqnarray}\label{eq:solutionTopInnerSurface}
&&\left(\begin{array}{c}u_{k}(\phi)\\v_{k}(\phi)
\end{array}\right)=\nonumber\\
&&c_{1,L}\left(\begin{array}{c}
-(\eps+\eta_z)D_{\frac{1}{2}(\eps^2-\eta^2_z)-1}\left(\sqrt{2}(\alpha\tan\phi-\kappa)\right)\\
\sqrt{2}D_{\frac{1}{2}(\eps^2-\eta^2_z)}\left(\sqrt{2}(\alpha\tan\phi-\kappa)\right)
\end{array}\right).
\end{eqnarray}

\subsection{Outer vertical surface ($\phi_0<\phi<\pi-\phi_0$)}
$r(\phi)=b/\sin\phi$
\begin{eqnarray}
ds^2&=&dx_1^2+dx_2^2=dx^2+\left(d b\cot\phi\right)^2\\
\Rightarrow x_1&=&x,\;\;\; x_2=a+b-b\cot\phi.
\end{eqnarray}
\begin{eqnarray}
A_1&=&Bb,\;\; A_2=0.
\end{eqnarray}
The physical spin operators (up to $\hbar/2$) are
\begin{eqnarray}
s_1&=&\sigma_3,\;\;s_2=\sigma_2,\;\; s_3=-\sigma_1.
\end{eqnarray}

For $\phi_0<\phi<\pi-\phi_0$ the Hamiltonian is
\begin{eqnarray}
H&=&\hbar v_F
\left[\sigma_1\left(\frac{1}{i}\frac{\partial}{\partial
x}-\frac{eB}{\hbar
c}b\right)-\sigma_2\frac{1}{i}\frac{\partial}{\partial
b\cot\phi}\right]-h_z\sigma_1.\nonumber\\
\end{eqnarray}
and the eigenfunctions are
\begin{eqnarray}\label{eq:solutionOuterSurface}
\left(\begin{array}{c}u_{k}(\phi)\\v_{k}(\phi)
\end{array}\right)&=&c_{2,1}e^{i\sqrt{\eps^2-(\kappa-\beta-\eta_z)^2}\beta\cot\phi}\left(\begin{array}{c}\alpha_+\\
\eps
\end{array}\right)\nonumber\\
&+&c_{2,2}e^{-i\sqrt{\eps^2-(\kappa-\beta-\eta_z)^2}\beta\cot\phi}\left(\begin{array}{c}\alpha_-\\
\eps
\end{array}\right),
\end{eqnarray}
where
\begin{eqnarray}
\alpha_{\pm}=\kappa-\beta-\eta_z\pm
i\sqrt{\eps^2-(\kappa-\beta-\eta_z)^2}.
\end{eqnarray}

\subsection{Bottom horizontal surface ($\pi-\phi_0<\phi<\pi+\phi_0$)}
$r(\phi)=-a/\cos\phi$

\begin{eqnarray}
ds^2&=&dx_1^2+dx_2^2=dx^2+\left(d a\tan\phi\right)^2\\
\Rightarrow x_1&=&x,\;\;\; x_2=2(a+b)+a\tan\phi.
\end{eqnarray}
\begin{eqnarray}
A_1&=&-Ba\tan\phi,\;\; A_2=0.
\end{eqnarray}

\begin{eqnarray}
s_1&=&\sigma_2,\;\; s_2=-\sigma_1,\;\; s_3=\sigma_3.
\end{eqnarray}

For $\pi-\phi_0<\phi<\pi+\phi_0$ the Hamiltonian is
\begin{eqnarray}
H&=&\hbar v_F
\left[\sigma_1\left(\frac{1}{i}\frac{\partial}{\partial
x}+\frac{a}{\ell^2_B}\tan\phi\right)+\sigma_2\frac{1}{i}\frac{\partial}{\partial
a\tan\phi}\right]\nonumber\\
&+&h_z\sigma_3+V_g.
\end{eqnarray}
where we included a different electrical potential on the bottom
surface $V_g$.

Near the outer vertical surface the solution on the bottom
horizontal surface is
\begin{eqnarray}\label{eq:solutionBottomOuterSurface}
&&\left(\begin{array}{c}u_{k}(\phi)\\v_{k}(\phi)
\end{array}\right)=\\
&&c_{3,R}\left(\begin{array}{c}\sqrt{2}D_{\frac{1}{2}((\eps-\nu_g)^2-\eta_z^2)}\left(\sqrt{2}\left(\kappa+\alpha\tan\phi\right)\right)\\
(\eps-\eta_z-\nu_g)
D_{\frac{1}{2}((\eps-\nu_g)^2-\eta_z^2)-1}\left(\sqrt{2}\left(\kappa+\alpha\tan\phi\right)\right)
\end{array}\right),\nonumber
\end{eqnarray}
where $\nu_g=V_g/(\hbar v_F/\ell_B)$.

Near the inner vertical surface the solution on the bottom
horizontal surface is
\begin{eqnarray}\label{eq:solutionBottomOuterSurface}
&&\left(\begin{array}{c}u_{k}(\phi)\\v_{k}(\phi)
\end{array}\right)=\\
&&c_{3,L}\left(\begin{array}{c}\sqrt{2}D_{\frac{1}{2}\left((\eps-\nu_g)^2-\eta^2_z\right)}\left(-\sqrt{2}(\kappa+\alpha\tan\phi)\right)\\
(\nu_g-\eps+\eta_z)D_{\frac{1}{2}\left((\eps-\nu_g)^2-\eta^2_z\right)-1}\left(-\sqrt{2}(\kappa+\alpha\tan\phi)\right)
\end{array}\right).\nonumber
\end{eqnarray}

\subsection{Inner vertical surface ($\pi+\phi_0<\phi<2\pi-\phi_0$)}
$r(\phi)=-b/\sin\phi$
\begin{eqnarray}
ds^2&=&dx_1^2+dx_2^2=dx^2+\left(d b\cot\phi\right)^2\\
\Rightarrow x_1&=&x,\;\;\; x_2=3(a+b)-b\cot\phi.
\end{eqnarray}
\begin{eqnarray}
A_1&=&-Bb,\;\; A_2=0.
\end{eqnarray}
\begin{eqnarray}
s_1&=&-\sigma_3,\;\; s_2=\sigma_2,\;\; s_3=\sigma_1.
\end{eqnarray}

For $\pi+\phi_0<\phi<2\pi-\phi_0$ the Hamiltonian is
\begin{eqnarray}
H&=&\hbar v_F
\left[\sigma_1\left(\frac{1}{i}\frac{\partial}{\partial
x}+\frac{b}{\ell^2_B}\right)-\sigma_2\frac{1}{i}\frac{\partial}{\partial
b\cot\phi}\right]+h_z\sigma_1.\nonumber\\
\end{eqnarray}
The eigenfunctions are
\begin{eqnarray}\label{eq:solutionInnerSurface}
\left(\begin{array}{c}u_{k}(\phi)\\v_{k}(\phi)
\end{array}\right)&=&c_{4,1}e^{i\sqrt{\eps^2-(\kappa+\beta+\eta_z)^2}\beta\cot\phi}\left(\begin{array}{c}\alpha_+\\
\eps
\end{array}\right)\nonumber\\
&+&c_{4,2}e^{-i\sqrt{\eps^2-(\kappa+\beta+\eta_z)^2}\beta\cot\phi}\left(\begin{array}{c}\alpha_-\\
\eps
\end{array}\right),
\end{eqnarray}
where
\begin{eqnarray}
\alpha_{\pm}=\kappa+\beta+\eta_z\pm
i\sqrt{\eps^2-(\kappa+\beta+\eta_z)^2}.
\end{eqnarray}

\subsection{Matching conditions}
As discussed in the main text, we require the continuity of the
wavefunctions near the outer surface where $\kappa\approx \beta$.
Therefore, we must have
\begin{eqnarray}
\left(\begin{array}{c}u_{k}(\phi^{(-)}_0)\\v_{k}(\phi^{(-)}_0)
\end{array}\right)&=&
\left(\begin{array}{c}u_{k}(\phi^{(+)}_0)\\v_{k}(\phi^{(+)}_0)
\end{array}\right)\\
\left(\begin{array}{c}u_{k}(\pi-\phi^{(+)}_0)\\v_{k}(\pi-\phi^{(+)}_0)
\end{array}\right)&=&
\left(\begin{array}{c}u_{k}(\pi-\phi^{(-)}_0)\\v_{k}(\pi-\phi^{(-)}_0)
\end{array}\right),
\end{eqnarray}
where $\phi^{(\pm)_0}=\phi_0\pm0^{+}$.

Using the wavefunctions determined in the Appendix, the above set of
four linear equations in four unknowns translates into
\begin{eqnarray}\label{eq:matchingOuterSurface}
\mathcal{A}_{+}c_{1,R}&=&e^{i\theta_{-}\alpha}(\omega_{-}+i\theta_{-})c_{2,1}+e^{-i\theta_{-}\alpha}(\omega_{-}-i\theta_{-})c_{2,2}\nonumber\\
\mathcal{B}_{+}c_{1,R}&=&e^{i\theta_{-}\alpha}\eps c_{2,1}+e^{-i\theta_{-}\alpha}\eps c_{2,2}\nonumber\\
\mathcal{B}_{g+}c_{3,R}&=&e^{-i\theta_{-}\alpha}(\omega_{-}+i\theta_{-})c_{2,1}+e^{i\theta_{-}\alpha}(\omega_{-}-i\theta_{-})c_{2,2}\nonumber\\
\mathcal{A}_{g+}c_{3,R}&=&e^{i\theta_{-}\alpha}\eps
c_{2,1}+e^{-i\theta_{-}\alpha}\eps c_{2,2}
\end{eqnarray}
where
\begin{eqnarray}
\mathcal{A}_{\pm}&=&\pm(\eps+\eta_z)D_{\frac{1}{2}(\eps^2-\eta^2_z)-1}\left(\sqrt{2}(-\beta\pm\kappa)\right)\\
\mathcal{B}_{\pm}&=&\sqrt{2}D_{\frac{1}{2}(\eps^2-\eta^2_z)}\left(\sqrt{2}(-\beta\pm\kappa)\right)\\
\mathcal{A}_{g\pm}&=&\pm(\eps-\nu_g-\eta_z)D_{\frac{1}{2}((\eps-\nu_g)^2-\eta^2_z)-1}\left(\sqrt{2}(-\beta\pm\kappa)\right)\nonumber\\
\\
\mathcal{B}_{g\pm}&=&\sqrt{2}D_{\frac{1}{2}((\eps-\nu_g)^2-\eta^2_z)}\left(\sqrt{2}(-\beta\pm\kappa)\right)\\
\omega_{\pm}&=&\kappa\pm\beta\pm\eta_z\\
\theta_{\pm}&=&\sqrt{\eps^2-\omega^2_{\pm}}.
\end{eqnarray}

Near the inner surface, where $\kappa\approx -\beta$, we must have
\begin{eqnarray}
\left(\begin{array}{c}u_{k}(\pi+\phi^{(-)}_0)\\v_{k}(\pi+\phi^{(-)}_0)
\end{array}\right)&=&
\left(\begin{array}{c}u_{k}(\pi+\phi^{(+)}_0)\\v_{k}(\pi+\phi^{(+)}_0)
\end{array}\right)\\
\left(\begin{array}{c}u_{k}(2\pi-\phi^{(+)}_0)\\v_{k}(2\pi-\phi^{(+)}_0)
\end{array}\right)&=&
\left(\begin{array}{c}u_{k}(-\phi^{(-)}_0)\\v_{k}(-\phi^{(-)}_0)
\end{array}\right),
\end{eqnarray}

This translates to
\begin{eqnarray}\label{eq:matchingInnerSurface}
\mathcal{B}_{g-}c_{3,L}&=&e^{i\theta_{+}\alpha}(\omega_{+}+i\theta_{+})c_{4,1}+e^{-i\theta_{+}\alpha}(\omega_{+}-i\theta_{+})c_{4,2}\nonumber\\
\mathcal{A}_{g-}c_{3,L}&=&e^{i\theta_{+}\alpha}\eps c_{4,1}+e^{-i\theta_{+}\alpha}\eps c_{4,2}\nonumber\\
\mathcal{A}_{-}c_{1,L}&=&e^{-i\theta_{+}\alpha}(\omega_{+}+i\theta_{+})c_{4,1}+e^{i\theta_{+}\alpha}(\omega_{+}-i\theta_{+})c_{4,2}\nonumber\\
\mathcal{B}_{-}c_{1,L}&=&e^{-i\theta_{+}\alpha}\eps c_{4,1}+e^{i\theta_{+}\alpha}\eps c_{4,2}\nonumber\\
\end{eqnarray}

\bibliography{tiHall}
\end{document}